\documentclass[10pt,a4paper,draft]{article}
\usepackage[latin1]{inputenc}
\usepackage{amsmath}
\usepackage{amsfonts}
\usepackage{amssymb}
\usepackage{graphicx}
\author{ Matej Hud\'{a}k \\ Stierova 23, SK-040 23 Ko\v{s}ice, Slovak Republic \\ 
hudakm@mail.pvt.sk \\ \\
Ondrej Hud\'{a}k\footnote{Corresponding author.}\\Faculty of Aerodynamics, Department of Aviation Technical Studies\\ Technical University Ko\v{s}ice\\ Rampova 7, SK 040 01 Ko\v{s}ice, Slovak Republic\\ hudako@mail.pvt.sk} 

\title{{\bf Multirelaxational Dynamic Response In Incommensurate Phases.}}

\begin{document}
\maketitle

\section*{Abstract}
Interactions of the uniform mode with higher order modes due to incommensurately modulated
equilibrium state change the usual relaxation behaviour to a more complex one.
A multirelaxation character should be present even in the single-plane-wave limit.
Our model enables one to describe dynamic dielectric response of the incommensurable
modulated phases in order-disorder systems.

\section{Introduction}

In this paper multirelaxational dynamic response in incommensurate phases is studied and extends results of our papers \cite{1}-\cite{5}. In \cite{6} Lovesey et al. studied response of the incommensurate (INC) phase. They found that it may become very complex at low frequencies : there are quasi particle-like excitations and dispersion-less bands of excitations. The origin of such response is in the fragmentation of the excitation energy spectrum. It occurs in magnetic materials, linear chains of atoms, electrons in the perpendicular magnetic field on lattice. In the incommensurate phase it is found that response exists at very low frequencies \( \omega \rightarrow 0 \) in spite of presence of strong site anisotropy.

In the ferroelectric material with an INC phase and dielectric relaxation (\(NaNO_{2} \) 
which has body-centered orthorhombic phase there is between \( T_{c} = 438.6 K \) and \( T_{f} = 437.4 K : \) a stable modulated phase. The modulation wave-vector varies with T from \( \frac{1}{8} \) at \( T_{c} \) to  \( \frac{1}{10} \) at \( T_{f} \) decreasing temperature. There is no evidence of commensurate plateaux. Orientational probability
of  the \( NO_{2} \) dipoles leads to a simple sine wave in the whole region of the modulated phase. Below \( T_{c} \) one expects  two relaxation mechanisms
with frequency independent relaxation times corresponding to the amplitude and phase
fluctuations, see in \cite{7}.

For a system of Ising spins close but above  \( T_{c} \) the order parameter dynamic susceptibility has a single Debye relaxation
\[ \chi({\bf q}, \omega)= \frac{\chi({\bf q},0)}{1+i \omega \tau({\bf q})}.\]

A monodispersive dielectric relaxation in the paraelectric phase in \( NaNO_{2} \) is found found by 
Hatta \cite{8}. The Cole-Cole diagram gives \( \beta=0.94 \) in the single Debye relaxation process. Hatta verified the validity of the single Debye-relaxation
model above \(T_{c}\):
\[ \tau_{para}= 2.4x10^{-8}(T-T_{0})^{-1.3}s. \]
\( T_{0}\) is the same as found from the fit of the static susceptibility above \( T_{c} \) to the Curie-Weiss law.
Below \( T_{c} \) an attempt to fit relaxation time to a single
effective relaxation time clearly indicates frequency dependence.
While the temperature behaviour of the effective relaxation time
resembles  that of the phase relaxations there is no reason
why phase relaxation should dominate the dielectric response at \( {\bf q}=0 \)
of the modulated structure. The frequency dependence of \( Im \chi \) in the temperature region
of the modulated phase shows clearly a shift of the weight of \( Im \chi \) with decreasing temperature to lower frequencies. As Hatta noted below \( T_{c}-0.5 \) the distribution
of the relaxation times gradually broadens.

In \( RbH_{3}(SeO_{3})_{2}\) there exists the incommensurate-ferroelectric phase transition.
The Cole-Cole diagram gives:
\[ 0.9 \le \beta \le 1 \]
with experimental accuracy \( \delta \beta \simeq 0.1 \), see in \cite{9}.
This may indicate a slightly polydispersive behaviour

In \( K_{2}SeO_{4} \) and \( Rb_{2}ZnCl_{4} \) the polydispersive dielectric process has been 
found \cite{10}. Note that in  \( Rb_{2}ZnCl_{4} \) the Cole-Cole diagram shows additional dispersive process increasing its contribution at lower frequencies (about 22 MHz) and lower temperatures ( in \( K_{2}SeO_{4} \) it is so about 5 MHz). The dielectric polydispersion is related by the authors to non-linear systems having discommensurations
where various kinds of modes are expected to be considered.

In \( \{N(CH_{3})_{4}\}_{2}ZnCl_{4} \) \cite{11} the monodispersive dielectric process of the Debye type is observed in the incommensurate phase except for close
vicinity of the incommensurate-commensurate transition \( T_{c}+0.02 K \) -- \( T_{c}+0.5 K  \)
where \( \beta \) is within \( 0.8-0.92 \). However above \( T_{c}+0.5 K  \) a single relaxation time is found. There are two additional dispersions: above 1 MHz frequency range
the Debye relaxation is observed and at low frequencies another one. The latter one is pronounced as temperature is lowered to \( T_{c} \) but the  dispersion frequency is almost temperature independent. The former process is thought to be connected with
the motion of discommensurations while the latter one is not (observed
by means of the signal field dependence).

In  \( Rb_{2}ZnBr_{4} \) the incommensurate phase has  unusual temperature independence of the modulation wavevector \cite{12}-\cite{14}. Its behaviour is inconsistent with the commonly
accepted point of view: T dependence of the real part of the dielectric tensor is due to increasing distance between discommensurations, as f.e. in \( K_{2}SeO_{4}. \)

In \( (NH_{4})_{2}BeF_{4} \) the polydispersive processes are present in the incommensurate phase \cite{15}.
The complex dielectric constant of \( (NH_{4})_{2}BeF_{4} \) single crystals was measured in the frequency range from 0.6 to 300 MHz in the vicinity of the transition temperature, the relaxation frequency is about 
108 Hz at \(T_{c}\). This is the order-disorder type system and its Cole-Cole diagram gives Debye type polydispersion with the coefficient \( \beta=0.7 \). In the lower frequency range
below 6 MHz: the additional low frequency dispersion process is present relaxational time of which exceeds  that for the "domain like structure".

In \( RbH_{3}(SeO_{3})_{2}\) the relaxational time \( 5.10^{-10} s \) \\
at \( 0.5^{o} \) above
the incommensurate-ferroelectric phase transition temperature is present, which is  typical
for order-disorder systems. \( \beta \) in Cole-Cole diagrams is:
\[ 0.9 \le \beta \le 1 \]
within experimental accuracy \( \delta \beta \simeq 0.1 \) \cite{16}. According to the authors: this is nearly monodispersive character; to be due to discommensurations.

In general in \( A'A''BX_{4} \) compounds a simple lattice model with Ising spins is a basis 
for qualitative discussions of relaxational phenomena in many structural phase transitions including displacive modulated \cite{17}. Then one may expect that discussion of the polydispersive processes in order-disorder materials may become relevant in general also to these systems.

Interactions in these materials are of the type: homogeneous mode -- higher order modes
due to incommensurate modulated ground state, does this fact change usual relaxation behaviour
to more complicated ? Is in this case a multirelaxation response present even in the single plane wave
limit? Should be domain-like structures in incommensurate dielectrics near the lock-in
transition to the lower ferroelectric phase responsible for distribution of the relaxation times in
the incommensurate structures ? How to increase the number of types of materials in which
the fragmentation of the excitation energy spectrum may be confronted with
reality?

Main results of this paper are:
The complex dynamical susceptibility was calculated using the Bloch
equations. Perturbation and nonperturbation approach enabled us
to derive formulae for susceptibility which display
multirelaxational behaviour due to the coupling of the
homogeneous polar soft mode to higher-order modes. We obtained results of numerical studies of multirelaxation phenomena due to interaction of modes and their comparison with experimental behaviour
in some dielectric materials. We introduced an approximative model which gives explicitly nonperturbation
results. At the of the paper there is a discussion.

\section{Model}

A model Hamiltonian \( H_{T} \) which we will use is well known (description in f.e. \cite{18}
A similar model is used in the description of modulated structures in magnetism \cite{19}-\cite{25} and in
in dielectrics ( \( NaNO_{2} \) ) \cite{26}, and in the ANNNI model (f.e. \cite{27}):
\begin{equation}
\label{1}
H_{1} = - \frac{1}{2} \sum_{i,j} J_{ij} S^{z}_{i} S^{z}_{j}.
\end{equation}
Here \( S^{z}_{i} \) is pseudospin variables \( S^{z}_{i} = \pm \frac{1}{2} \), we assume 
a simple cubic lattice \( i = 1,...,N_{latt}, \) where \( N_{latt} \) is number of sites, each site is  a fluctuating unit (f.e. H, D, \( NO_{2} \) ).
In real materials (KDP, \( NaNO_{2} \) type, etc. ) there are more complicated structures which need for description more complex models.

We use for simplicity:
\[ Q \equiv {\bf Q.a}, \]
 where \( {\bf a} \) is a basic lattice vector (1,0,0). The interaction 
 energy J({\bf q}) for  {\bf q} in the {\bf a} direction is given by:

\[      J({\bf q}) = 2J_{1} \cos({\bf qa}) + 2J_{2} \cos(2 {\bf qa})
+ 4J_{perp}. \]

The random-phase-approximation type analysis gives:
\[ <S^{z}_{n}>= \frac{1}{2} \tanh( \beta \frac{H_{n}}{2}), \]
\[ H_{n} = \sum_{j} J_{nj}<S^{z}_{j}>, \]
and  \( kT_{c} = \frac{J(Q)}{4} \) defines the transition temperature \(T_{c})\).
The single plane wave modulated state is given by:
\begin{equation}
\label{5}
<S^{z}_{n}>=  S \cos( Qn + \phi ).
\end{equation}
Here \( \phi \) is an arbitrary phase and the local field on the site \()n\) is:
\[  H_{n} = SJ(Q) \cos(Qn + \phi ).  \]

The modulation vector amplitude is:
\begin{equation}
\label{6}
\cos(Q) = -\frac{J_{1}}{4J_{2}}.
\end{equation}

Note that Q is incommensurate if it is different from the number \( 2 \pi \frac{M}{N} \),
where M, N are any integers.

In real materials we have: for DRADP: \( \frac{Q}{2 \pi} \approx 0.35 \), for \( Rb_{2}ZnBr_{4} \):  \( \frac{Q}{2 \pi}  \approx \frac{5}{17} \), for \( NaNO_{2} \):  \( \frac{Q}{2 \pi} \approx \frac{1}{8} \)
immediately below \( T_{c} \).

The equilibrium state energy:
\begin{equation}
\label{7}
E_{GS} = - N_{latt} J(Q) \frac{S^{2}}{2},
\end{equation}
for small amplitude of the basic harmonics, S, given by:
\begin{equation}
\label{8}
S^{2} \equiv \frac{T^{3}}{T^{3}_{c}}( \frac{T_{c}}{T} - 1).
\end{equation}

Numerical values of model constants for \(NaNO_{2} \) are: \( J_{1} \approx 81 K \), 
\(  J_{2} \approx -26 K \), \( J_{perp} \approx 410 K \), \( T_{c}=438.69 K. \)
The Curie-Weiss temperature for the direct para-ferroelectric (virtual) transition is \( T_{0}=437.41 K \),
here the virtual transition temperature is \( kT_{0} = \frac{J(0)}{4}. \)
However the transition temperature for the incommensurate to
ferroelectric phase transition is 436.24 K. The phase above the incommensurate one is the paraphase.

The high temperature relaxation time constant estimated is estimated to be \( T_{1}=0.7922 K^{-1} \).
The incommensurate parameter is \( \delta = \frac{1}{8} \).

The small amplitude approximation in (\ref{8}) as restricted to
temperatures not very low with respect to the transition temperature is:
\[S^{2} \approx ( 1- \frac{T}{T_{c}}).\]

Let us assume that Q  is a modulation vector amplitude in the incommensurate single plane wave
modulated phase. The dynamics of excitations using Bloch
equations for the motion of pseudospins \cite{1} may be described by the method well known and thoroughly discussed in \cite{18} in which the equations of motion for pseudospins is:
\begin{eqnarray}
\label{17}
i \omega \delta <S_{i}^{z}> & = & - \frac{1}{T_{1}} [ \delta <S_{i}^{z}>
- \frac{ \beta }{4} ( 1 - 4 <S_{i}^{z}>^{2} ). \nonumber \\
&   & . \sum_{j} J_{ij} \delta <S_{j}^{z}> - \frac{ \beta }{2}
( 1 - 4 <S_{i}^{z}>^{2} ) \mu E_{i} ],
\end{eqnarray}
where
\(  \delta <S_{i}^{z}> \) is  mean values of amplitudes of fluctuating pseudospin at site i
\[ <S^{z}_{i}>_{t} = <S^{z}_{i}> + \delta <S_{i}^{z}> \exp(i\omega t), \]
here \( \mu \) is the dipole moment of the elementary unit, \( T_{1} \) is the longitudinal high-temperature relaxation time, the molecular field \( {\bf H}_{i} \) at site i:
\begin{equation}
\label{18}
{\bf H}_{i} = ( 0, 0, \sum_{j} J_{ij} <S^{z}_{j}> ) \equiv ( 0, 0, H_{i} ).
\end{equation}

If a general time and site dependent electric field
\( E_{i}(t) \) is applied to the equilibrium state (\ref{5}) then an infinite set of equations (the Fourier transform of (\ref{17})) is given by:
\begin{equation}
\label{19}
a_{n} S_{n} + b_{n-1} S_{n-1} + b_{n+1} S_{n+1} = c e_{n} + g e_{n-1} + g e_{n+1},
\end{equation}
where n is integer, here \( \beta \equiv \frac{1}{k_{B}T}, \) and
\begin{eqnarray}
\label{20}
a_{n} & = & ( i \omega + \frac{1}{T_{1}} - \frac{ \beta J(q+2nQ)}{4T_{1}}
(1-2S^{2}))   \\
b_{n} & = & \frac{ \beta S^{2}}{4T_{1}} J(q+2nQ) \nonumber \\
c & = & \frac{ \beta \mu }{2 T_{1}} (1-2S^{2}) \nonumber \\
g & = & - \frac{ \beta \mu }{2 T_{1}} S^{2} \nonumber \\
E_{n} & = & E_{q+2nQ} \nonumber \\
e_{n} & = & E_{n} \exp(-i2 \phi n ) \nonumber \\
S_{n} & = &  \delta<S^{z}_{q+2nQ}> \exp(-i2 \phi n ) \nonumber .
\end{eqnarray}

Note that the homogeneous field is coupled to an infinity
set of modes in the incommensurate state.

The case of a commensurate phase: a periodicity in the n variable is found and a finite number of quasiparticle bands. The incommensurate case fluctuation spectrum is complex in
the exact mathematical sense - a well known Cantor set of points in the
energy spectrum.

\section{Paraphase}

In the paraphase the susceptibility can be calculated from:
\[ \chi = \frac{2\mu \sum_{i} \delta <S_{i}^{z}>}{N_{latt} E} \]
in a homogeneous electric field with amplitude E. We obtain:
\begin{equation}
\label{21.1}
\chi_{para} = \frac{ \frac{ \mu^{2} }{T_{1} k_{B} T}}
{i \omega + \frac{1}{ T_{1} } (1 - \frac{J(0)}{4 k_{B} T})}.
\end{equation}
Here we find the Debye relaxation type behaviour:
\[  \frac{1}{T_{1}} (1 - \frac{T_{0}}{T_{c}}) \] at \( T_{c}. \)

where for \( NaNO_{2} \) \( J(O) > 0 \), this behaviour observed by Hatta \cite{28}-\cite{29}.

\section{Incommensurate phase}

Below \( T_{c} \), the transition temperature to the incommensurate phase, S is the perturbation parameter
we calculate when neglecting coupling of the \( q = 0 \) mode to other modes
(would be absent in the uniform equilibrium state). Then the dielectric susceptibility:
\begin{equation}
\label{21.2}
\chi_{INC} =
\frac{ \frac{ \mu^{2} (1 - 2 S^{2})}{ T_{1} k_{B} T}}
{ i \omega + \frac{1}{T_{1}} - \frac{ \beta J(0)}{4T_{1}}
(1-2S^{2})}.
\end{equation}

We find the Debye-relaxation type dispersion, the C constant in \( \chi_{INC} \) is modified
with respect to its paraphase value by the factor \( (1-2S^{2}) \). Note: for the uniform phase (with Q=0) 2 in this factor should be replaced by 4.

If modulation wave vector Q is commensurate then a finite number of modes \( S_{0, \pm 1, ... } \)
would contribute to the susceptibility. If Q is incommensurate theoretically all higher order terms
will contribute to the susceptibility. For the dielectric susceptibility of the incommensurate phase up to
the fourth order in S we find:
\begin{eqnarray}
\label{21.4}
\frac{\chi_{4}}{2 \mu} & =  & \frac{\chi_{para}}{2 \mu}
-S^{2} ( \frac{c_{1}}{\alpha_{0}} + \frac{c_{0} \beta_{0}}{\alpha_{0}^{2}})
+S^{4} ( \frac{c_{0} \beta_{0}^{2}}{\alpha_{0}^{3}})- \nonumber \\
 &  & - \frac{g b_{1}}{\alpha_{0} \alpha_{1}}
- \frac{g b_{-1} }{\alpha_{0} \alpha_{-1} }
+ \frac{c_{0} b_{0} b_{1}}{\alpha_{0}^{2} \alpha_{1}}
+ \frac{c_{0} b_{0} b_{-1}}{\alpha_{0}^{2} \alpha_{-1}} + O(S^{6}),
\end{eqnarray}

where
\begin{eqnarray}
\alpha_{n} & = & ( i \omega + \frac{1}{T_{1}} - \frac{ \beta J(q+2nQ)}{4T_{1}}
)  \nonumber \\
\beta_{n} & = &  \frac{ \beta J(q+2nQ)}{2T_{1}} \nonumber \\
a_{n} & = & \alpha_{n} + S^{2} \beta_{n} \nonumber \\
c_{0} & = & \frac{ \beta \mu }{2 T_{1}}  \nonumber \\
c_{1} & = & \frac{ \beta \mu }{ T_{1}}  \nonumber \\
c & = & c_{0} - c_{1} S^{2}. \nonumber \\
\end{eqnarray}

The expansion in powers of \( S^{2} \) is valid when the following condition is fulfilled
\[ \mid i \omega + \frac{1}{T_{1}}(1-\frac{J(q+2nQ)}{4k_{B}T}) \mid >>
\mid \frac{\beta S^{2}J(q+2nQ)}{2T_{1}} \mid. \]
This condition is satisfied for sufficiently large frequencies,
which are still smaller than \( T_{1} \).
At very low frequencies another approach should be used.

The frequency dependence of the real and imaginary part of the dielectric
susceptibility calculated for both cases when the mode coupling is neglected
(1) and when the mode coupling is taken into account (2) up to the fourth
order in S. The contribution to \( Im \chi \) due to fourth order terms gives higher
absorption peak, an increase of this quantity in the low frequency tail, tendency to create asymmetry of the peak present in the paraphase on the low frequency side.

The difference between \( Im \chi \) calculated with and without coupling is in the fact that 
the coupling contribution has a peak-like shape and a tail to low frequencies.
The peak maximum is at lower frequency than the peak maximum of \( Im \chi \).

The difference between the real part of the susceptibility \( Re \chi \) calculated
with and without coupling gives that changes are most pronounced at low frequencies.

Let us note that general time symmetry arguments gives that real and imaginary part of the susceptibility
into the Debye-like form:
\[ \chi = \frac{C(\omega)}{i \omega+ \frac{1}{\tau(\omega)}}. \]
the effective (real) relaxation time
\[ \frac{1}{\tau_{eff}(\omega)} = \frac{\omega
\chi^{,}}{\chi^{,,}}, \]
the effective (real) constant \( C(\omega) \)
\[ C(\omega) =
\frac{\omega((\chi^{,})^{2}+(\chi^{,,})^{2})}{\chi^{,,}}. \]

There is a temperature dependence of the effective relaxation frequency \( \tau^{-1}_{eff} \) for
\(NaNO_{2} \) for \( \delta = \frac{1}{8}\) at low frequencies.
Frequency corresponding to the curve (1) is 2.5 times lower than the frequency corresponding to the curve (2).

The same frequency dependence was observed by Hatta (above) and discussed in \cite{7}.
Note that existing measurements are done for  a few different frequencies only.
Desirable is to perform systematic measurements for more frequencies to  understand
how temperature dependence of the relaxation time varies with frequency, frequency dependence of the
effective relaxation frequency \( \tau^{-1}_{eff} \),  temperature and frequency
dependence of the effective constant C, temperature and frequency dependence of the
effective relaxation frequency  and  frequency dependence of the effective constant C
originate in the multirelaxation processes. Both quantities are frequency independent in the single Debye
relaxation process. Between low frequency and high frequency we find that above there are fast relaxation processes, and below there are relaxation processes slower (effective relaxation frequency is lower).
The different behaviour  may be explained by different relaxation times of processes which are effective
above and below this region. Expectation that the coupling between the homogeneous
mode and higher order modes influences mainly the low frequency dynamics should be verified by nonperturbation calculations.

\section{Mode-independent approximations}

Purpose of this section is to introduce several approximations within previously defined
model, which enable to study equations (\ref{20}) in a mathematically closed form.
We find that the solution is physically  non-trivial, displays new features concerning multirelaxational response of incommensurable modulated phases.

The paraphase - ferrophase phase transition usually is splitted into at least two transitions:
paraphase - incommensurate phase and incommensurate phase - ferrophase.
INC phase temperature region is small near virtual phase transition paraphase - ferrophase temperature.

The interaction energy J(q) has a maximum value at q=Q -- modulated state, and a minimum value at q=0 -- ferroelectric state ( \( NaNO_{2} \))
\begin{equation}
\label{Jd}
\delta  \equiv \frac{J(Q) - J(0)}{J(0)} = \frac{T_{c}-T_{0}}{T_{0}}.
\end{equation}

Consider systems in which (max) value of J(Q) is only slightly higher than lowest J(0) value.
Estimate for \( NaNO_{2} \) gives 0.005.

J(q) for any but q=Q is approximated in the zero order in \( \frac{T_{c}-T_{0}}{T_{0}}\) by:
\[ J(q) \approx J(0)+ \frac{J(Q)-J(0)}{2}(\delta_{q,Q}+\delta_{q,-Q}). \]
Note: this approximation gives simplified dynamic equations. J(q) in direct space has the form:
\[ J(r-m) = J(0) \delta_{r,m} +(J(Q)-J(0)) \cos(Q(r-m)). \]
J(r) is infinity-range effective oscillatory pseudospin-pseudospin interaction energy.
Single wave ground state remains still sinusoidally modulated with q=Q.
Now we have:
\[ J(2nQ)= J(0) \]
for all non-zero integers n if Q takes an incommensurate value.
Thus we assume: for those values of the system parameters, for which the quantity in (\ref{Jd}) is
small, the incommensurate phase may be practically characterized by the single plane wave regime.
We neglect quantities of the order \( \frac{T_{c}-T_{0}}{T_{0}} \) and higher,
while we still consider fluctuations \(S_{n} \) of the order \( S^{\vert 2n \vert}. \)
It is clear that our approach is more correct for temperatures
nearby the transition from paraphase to incommensurate phase.

Mode-independent approximation gives mode independent relaxation time:
\[  \frac{1}{T_{1}} - \frac{ \beta J(q+2nQ)}{4T_{1}}(1-2S^{2}), \]
an effective interaction energy J(0) leads to a single relaxation frequency
\[  \frac{1}{T_{1}} - \frac{ \beta J(0)}{4T_{1}}(1-2S^{2}). \]
\( a_{n} \) from (\ref{20}) take the same form for each mode:
\[ a_{n} =  ( i \omega + \frac{1}{T_{1}} - \frac{ \beta J(0)}{4T_{1}}
(1-2S^{2})) \equiv (i \omega + a), \]
which is a mode (index n) independent form.

Physics in this approximation is that an original set of different relaxation times is represented by
the largest one. Mode independent relaxation energy approximation gives in (\ref{20}) a set of interaction constants
\[ b_{n}  =  \frac{ \beta S^{2}}{4T_{1}} J(q+2nQ), \]
for different modes of fluctuations. An effective interaction energy J(0):
\[ b_{n}  =  \frac{ \beta S^{2}}{4T_{1}} J(0) \equiv b, \]
which is mode (index n) independent form.

An approximation for the dynamic equations assuming mode independent approximation to the equations (\ref{20}) is:
\begin{equation}
\label{19'}
(i \omega +a) S^{0}_{n} + b S^{0}_{n-1} + b S^{0}_{n+1} = c e_{n} + g e_{n-1} +
g e_{n+1}.
\end{equation}
Here \( \delta a_{n} = a_{n} -a \), \( \delta b_{n} = b_{n} -b \), \( \delta S_{n} = S_{n} -S^{0}_{n} \).
\(S_{n} \) is assumed to satisfy (\ref{20}). (\ref{19'}) is equivalent to the original
equations (\ref{20}) if we simultaneously  solve also:
\begin{equation}
\label{19''}
a_{n} \delta S{n} + b_{n-1} \delta  S_{n-1} + b_{n+1} S_{n+1} =
- \delta a_{n} S^{0}_{n} - b_{n-1} S^{0}_{n-1} - b_{n+1} S^{0}_{n+1}.
\end{equation}
Thus our approximation to reduce (\ref{20}) to (\ref{19'})
is the same as to (\ref{19''}) with respect to (\ref{19'}).
Further we discuss only (\ref{19'}).

An approach  to incomensurability via a limiting process using a sequence of commensurate phases:
the infinite set (\ref{19'}) easily solved using limiting procedure: 
the incommensurate phase with the modulation wavevector Q is approached by a sequence of
commensurate phases with the modulation wavevectors Q(L,M)
\[ Q(L,M) \equiv \frac{2 \pi M}{L}, \]
where
\[ \lim_{M,L \rightarrow \infty} Q(L,M) = Q. \]

Steps in the procedure are: a procedure which leads to solution of (\ref{20})
consists of solution of (\ref{19'}) for finite appropriate L and M integers,
and then taking the limit above

A solution for a general mode is, assuming that Q is commensurate (M and L), in the form
\( S_{n} \):
\[ S_{n} = \sum_{\kappa = \frac{-L}{2}+1}^{\frac{L}{2}} \exp{ (i
\frac{2\pi}{L}\kappa n)} S_{\kappa}, \]
for any value of momentum q. Note that the periodicity condition holds \( S_{n+L} = S_{n} \).
From (\ref{19'}) we obtain
\[ S_{\kappa} =    \frac{1}{L} \sum_{n} \frac{(c+ge_{n-1}
+ge_{n+1})
\exp{(- i \frac{2\pi}{L} \kappa n)}}{i \omega + a +2b \cos(\frac{2\pi
\kappa}{L})}. \]
The inverse form, the homogeneous field has the form:
\begin{equation}
\label{sn}
S_{n} =  \frac{1}{L} E \sum_{\kappa} \frac{  (c + 2g \cos(\frac{2 \pi
\kappa}{L}))}{i \omega + a +2b \cos(\frac{2\pi
\kappa}{L})}  \exp{ (i \frac{2\pi}{L}\kappa n)},
\end{equation}
which is a general form for the motion of the n-th mode due to the external field.

\section{Dynamic susceptibility}

Dynamic dielectric susceptibility \( \chi \) taking appropriate limit to Q is given by:
\begin{equation}
\label{ds}
\chi = \frac{\mu N_{latt}}{\pi} \int_{-\pi}^{+\pi}
\frac{  (c + 2g \cos(x))}{i \omega + a +2b \cos(x)} dx .
\end{equation}

The real part of the susceptibility is:
\begin{equation}
\label{dsr}
\chi^{'} = \frac{\mu N_{latt}}{\pi} \int_{-\pi}^{+\pi}
\frac{ (a +2b \cos(x)).(c + 2g \cos(x))}{ \omega^{2} + (a +2b \cos(x))^{2}}
dx .
\end{equation}

The imaginary part of susceptibility is:
\begin{equation}
\label{dsi}
\chi^{''} = - \frac{\omega \mu N_{latt}}{\pi} \int_{-\pi}^{+\pi}
\frac{ (c + 2g \cos(x))}{ \omega^{2} + (a +2b \cos(x))^{2}} dx.
\end{equation}

\section{Static susceptibility}

The zero-frequency limit of (\ref{ds}) is the static susceptibility \( \chi_{0} \):
\begin{equation}
\label{s0}
\chi_{0} = \frac{\mu^{2}N_{l}}{k_{B}T}
\frac{1}{\sqrt{\frac{T-T_{0}}{T}}
\sqrt{\frac{T-T_{0}}{T}+\frac{4S^{2}T_{0}}{T}}}
\end{equation}
\[ .[ 1-2S^{2}-\frac{T}{T_{0}}(\sqrt{\frac{T-T_{0}}{T}}
\sqrt{\frac{T-T_{0}}{T}+\frac{4S^{2}T_{0}}{T}})-
(1-\frac{T}{T_{0}}(1-2S^{2}))]. \]

Decreasing temperature, above \( T_{c} \), behaviour is of the mean field type:
\[ \chi_{0} \approx \frac{1}{T-T_{0}}. \]
The critical index in this temperature region is 1.
Decreasing further temperature below \( T_{c}, \) but still above
\( T_{0} \) gives that the critical behaviour of the static susceptibility
changes to
\[ \chi_{0} \approx \frac{1}{\sqrt{T-T_{0}}} \]
and the critical index changes to \(  \frac{1}{2} \).
Our model predicts different critical behaviour for the static susceptibility in the modulated region.
The behaviour of the inverse static susceptibility (\ref{s0}) has 
a local minimum around \( T_{c} \) which is then changed to a local maximum followed by a steep decrease,
This may describe uncertainty as concerning the experimental behaviour of
the dielectric  static susceptibility, see in \cite{7} where the values of the critical index is reported as 1.11-1.24. Those values are obtained from fits within the paramagnetic phase, where
dielectric static susceptibility displays departure from the mean
field behaviour, which occurs above 500 K only.
Thus there is a question: Do interacting fluctuations of the forming incommensurate phase renormalise
paraphase behaviour in the way corresponding to experiments? Do our results indicate such possibility?

\section{Dispersion-less modes}

The response of the incommensurate structure consists of a sum of infinite number
of relaxators with their relaxation frequencies:
\[ \frac{1}{\tau_{x}} \equiv a +2b \cos(x). \]
After substitution of a and b constants we obtain:
\[ \frac{1}{\tau_{x}} = \frac{T-T_{0}}{T_{1}T}+
\frac{T_{0}S^{2}\cos^{2}(\frac{x}{2})}{T_{1}T} \]
The effective temperature \( T_{0} \) is defined as
\[ \frac{J_{0}}{4k_{B}}. \]
The maximum of the relaxational frequency is:
\[ \frac{T-(1-S^{2}T_{0})}{T_{1}T}. \]
and the minimum of the relaxational frequency is:
\[ \frac{T-T_{0}}{T_{1}T}. \]
Both extremal values are the same when the temperature increases to the transition temperature \( T_{c} \).
Their difference reveal such a behaviour explicitly:
\[ \max \frac{1}{\tau_{eff}(x)}-\min \frac{1}{\tau_{eff}(x)} =
\frac{S^{2}T_{0}}{T_{1}T}. \]
Thus: spread of relaxational frequencies increases decreasing temperature below the transition temperature.
The Curie constants are:
\[  \frac{\mu N_{latt}}{\pi}   \frac{c + 2g \cos(x)}{a+2b\cos(x)}. \]
Both quantities, relaxation frequency and Curie constant,
are mode dependent, see their x-dependence.

Another form of the dynamic susceptibility is:
\begin{equation}
\label{ds1}
\chi = \chi_{0} \int_{-\pi}^{+\pi} \frac{w(x)}{i \omega
\tau_{x}+1} dx,
\end{equation}
here \( \chi_{0} \) is the static susceptibility, the weight of an x-relaxation mode contribution to the
susceptibility is:
\[ w(x) \equiv \frac{1}{2\pi \chi_{0}} \frac{1+ 2 \frac{g}{c}\cos(x)}{1+ 2
\frac{b}{a}\cos(x)}. \]
The weight is normalized to unity:
\[ 1 =  \int_{-\pi}^{+\pi} w(x) dx. \]
The explicit form of the weight function is:
\[ w(x) \equiv \frac{1}{2\pi \chi_{0}} \frac{1-  \frac{2S^{2}}{1-2S^{2}}\cos(x)}
{1+  \frac{2S^{2}}{2S^{2}-1+\frac{T}{T_{0}}}\cos(x)}. \]
Below the critical temperature and above the effective temperature \( T_{0} \) there exists
no singular point in the weight function.
Interpretation of the form (\ref{ds1}) is: susceptibility = the sum of an infinite number of relaxation
modes, each characterized by the label x, its density w(x), and
its relaxation frequency \( \frac{1}{\tau_{x}}\).
These modes contribute at the same frequency \(\omega\). We have found continuum of states
of excitations characterized by the continuously varying weight
w(x) instead of the quasiparticle-like spectrum (the weight function
of which consisting of a sum of singular delta terms).
An interpretation of excitation spectra within our model is closely  related to that for other
systems made by Lovesey and his coworkers \cite{22} and \cite{25}.
Lovesey's approach usually predicts also existence of nonzero density of states at zero frequency, 
our approach gives the quantity \( y_{-} > 0 \) as the lowest frequency of these excitations.

\section{Dielectric losses}

The imaginary part of the dynamic susceptibility as transformed is:
\begin{equation}
\label{drii}
\chi^{''}= - \frac{\chi_{0} \omega}{\pi} \int^{y_{+}}_{y_{-}} dy
\frac{1}{\sqrt{(y_{+}-y)(y-y_{-})}}
\frac{1+\frac{2g.(2y-\delta_{+})}{\delta_{-}}}{y^{2}+\omega^{2}}
\end{equation}
where
\[ \chi_{0} \equiv 2 \mu N_{latt} \frac{c}{a}, \]
\[ \delta_{\pm} \equiv y_{+} \pm y_{-} . \]
Here \( y = a+2b \cos(x). \) From (\ref{drii}): the weight w(y) is proportional to
the inverse square roots:
\[ w(y) \approx \frac{1}{\sqrt{(y_{+}-y)(y-y_{-})}}. \]
The first root has singularity at the point \( y_{+} \):
\[ y_{+}= \frac{1}{T_{1}} (1-\frac{\beta J(0)}{4k_{B}}(1-4S^{2})). \]
Note: it does not depend on temperature if we use:
\[ S^{2} = \frac{1}{4} (1-\frac{T}{T_{c}}) \]
instead of the original simplified one:
\[ S^{2} =  (1-\frac{T}{T_{c}}). \]
Then \( y_{+} \) takes the form:
\[ y_{+}= \frac{1}{T_{1}} (1-\frac{T_{0}}{T_{c}}). \]
The explicit form of the lower root singularity is:
\[ y_{-}= \frac{1}{T_{1}} (1-\frac{T_{0}}{T}) \]
Note: both quantities, \( y_{\pm} \) are positive for
temperatures higher than \( T_{0} \). Decreasing temperature
towards \( T_{0} \) the quantity  \( y_{-} \) vanishes
and lower and lower relaxation frequencies become active.
Real transition temperature from the incommensurate to the
ferroelectric phase is slightly higher than temperature \( T_{0}. \)
The weight of low frequency relaxation movements of pseudospins increases decreasing temperature.

\section{Real part of the susceptibility}

The real part of the dielectric susceptibility is easily transformed to the more instructive form:
\begin{equation}
\label{drrr}
\chi^{'}=  \frac{\chi_{0} }{\pi} \int^{y_{+}}_{y_{-}} dy
\frac{y}{\sqrt{(y_{+}-y)(y-y_{-})}}
\frac{1+\frac{2g.(2y-\delta_{+})}{\delta_{-}}}{y^{2}+\omega^{2}}.
\end{equation}
interpretation of which is easy.

\section{Frequency behaviour}

Low frequency behaviour for temperatures \( T > T_{0} \) is the following.
note that for \( T > T_{0} \) we have \( y_{-} > 0 \) and
the low frequency here is defined by the condition:
\[ \omega < y_{-}. \]
From (\ref{drii}) and (\ref{drrr}) we obtain approximative expressions for real and imaginary dielectric susceptibilities:
\[ \chi^{'} \approx A - B \omega^{2}+ O(\omega^{4}) \]
and
\[ \chi^{''} \approx C \omega - D \omega^{3}+ O(\omega^{5}). \]
Here we have found that:
\[ a \equiv 1- 2 \frac{g}{c} \frac{y_{+}+y_{-}}{y_{+}-y_{-}} \]
\[ b \equiv 4 \frac{g}{c} \frac{1}{y_{+}-y_{-}} \]
\[ A= I_{2} a + I_{1} b \]
\[ B= I_{4} a + I_{3} b \]
\[ C= I_{1} a + I_{0} b \]
\[ D= I_{3} a + I_{2} b \]
\[ I_{0} = 1 \]
\[ I_{1} = \frac{1}{\sqrt{y_{+}y_{-}}}\]
\[ I_{2} = \frac{y_{+}+y_{-}}{2 (y_{+}y_{-})^{\frac{3}{2}}}\]
\[ I_{3} = \frac{3(y_{+}+y_{-})^{2}-4y_{+}y_{-}}{8
(y_{+}y_{-})^{\frac{5}{2}}}\]
\[ I_{4} = \frac{15(y_{+}+y_{-})^{3}-36y_{+}y_{-}(y_{+}+y_{-})}
{48 (y_{+}y_{-})^{\frac{7}{2}}}. \]
Maximal losses are at the frequency:
\[ \omega_{e}^{2} \approx \frac{C}{3D}. \]
The halfwidth of this peak is:
\[ HW = \frac{2 \omega_{e}}{\sqrt{3}}. \]
The effective relaxational frequency becomes frequency dependent:
\[ \frac{1}{\tau_{eff}} \equiv \frac{\chi^{'} \omega}{\chi^{''}}
\approx \frac{1}{\tau_{eff}}(0) . (1+ \rho \omega^{2} ) \]
where
\[ \frac{1}{\tau_{eff}}(0) \equiv \frac{A}{C} \]
and
\[ \rho \equiv \frac{B}{A}- \frac{D}{C}. \]
The effective relaxational frequency quadratically increases
when frequency increases, at least at low frequencies and for
temperatures immediately below the transition temperature.

Calculating approx. constants A,B,C and D we obtain:
\begin{equation}
\label{tef}
\frac{1}{\tau_{eff}} \approx \frac{1}{\tau_{eff}}(0) . (1+
\alpha ( \omega^{2} \tau_{eff}^{2}(0)) S^{4}.
\end{equation}
The correction to zero-frequency behaviour is of the order \(S^{4} \) in agreement with our previous perturbation calculations. Zero frequency effective relaxational frequency \( \frac{1}{\tau_{eff}}(0)
\) is temperature dependent:
\[  \frac{1}{\tau_{eff}}(0) =
\frac{1-T_{1}\sqrt{\alpha^{2}-4b^{2}}}{\frac{\alpha}
{\alpha^{2}-4b^{2}}-T_{1}} \]
where
\[ \alpha \equiv  \frac{1}{T_{1}} - \frac{ \beta J(0)}{4T_{1}}
(1-2S^{2}) \]
\[ b \equiv  \frac{ \beta S^{2}}{4T_{1}} J(0) . \]
The effective zero-frequency relaxation frequency below transition temperature from paraphase to
the incommensurate phase is :
\[   \frac{1}{T_{1}} - \frac{ \beta J(0)}{4T_{1}}. \]
At the transition point both quantities are continuously connecting each other:
\[  \frac{1}{\tau_{eff}}(0) \approx \alpha \approx (T-T_{0}),\]
below this temperature point fluctuations renormalise the temperature dependence
\[  \frac{1}{\tau_{eff}}(0) \approx \alpha-2b \approx (T-T_{0}) .\]
Softening of the relaxational frequency below \( T_{c} \)
continues  from above this point, however proportionality
coefficient is renormalized.

Experiments by Hatta (1970) \cite{9} for \( NaNO_{2} \) reveal that
immediately below the transition to the incommensurate phase
frequency dependence of the effective relaxation frequency, behaviour found within this theory corresponds qualitatively with that observed by Hatta: increasing frequency relaxational frequency increases too.
Also temperature dependence qualitatively corresponds to observed. More precise comparison of
our predictions with experiments is desirable and not only for \( NaNO_{2} \) materials.

High frequency behaviour for temperatures \( T > T_{0} \) has \( y_{+} > 0 \),
which remains finite for considered temperature region. The high frequency region is defined by the condition:
\[ \omega > y_{+}. \]

Both parts of the susceptibility have approximate frequency behaviour:
\[ \chi^{'} \approx  F \frac{1}{\omega^{2}}+ O(\omega^{-4}) \]
and
\[ \chi^{''} \approx G \frac{1}{\omega}+ O(\omega^{-3}). \]
Here F and G are temperature dependent constants. Both parts of the susceptibility
vanish in power-like way increasing frequency.
Qualitatively such behaviour resembles behaviour of a single Debye-like relaxator.
The effective relaxational frequency becomes frequency independent:
\[ \frac{1}{\tau_{eff}} \equiv \frac{\chi^{'} \omega}{\chi^{''}}
\approx  \frac{F}{G} . \]

Intermediate frequency behaviour for temperatures  \( T > T_{0} \),
where the intermediate frequency is defined by the condition:
\[ y_{-} < \omega < y_{+}, \]
gives below the transition from para to incommensurate phase that the intermediate frequency region is very
small, decreasing temperature  its range increases.
When the inequality above holds, then both parts of the susceptibility cannot be
expanded directly in powers  of frequency.

\section{Discussion}
Let us first discuss our results for \( NaNO_{2} \) and then for other systems.

\subsection{ \( NaNO_{2} \) }

Within our results it is qualitatively possible to interpret temperature and
frequency dependence of the effective relaxation frequency observed by Hatta.
In high temperature  region a single relaxation process takes place, and the
relaxational frequency  is proportional to temperature difference \( T-T_{0} \).
In the temperature region immediately below the transition temperature \( T_{c} \)
a small local maximum in temperature dependence of the effective relaxation frequency occurs.
In the temperature region further below the transition
temperature \( T_{c} \) a decrease of the effective relaxation frequency with
decreasing temperature occurs. Below the transition temperature \( T_{c} \)
the effective relaxation frequency is frequency dependent,
its value increases with increasing frequency. It would be of interest to perform a systematic experimental
test of our theory, especially concerning the frequency dependence
of the effective relaxation frequency.

\subsection{Other systems}

A simple lattice model with two sets of coupled Ising spins may explain many structural
phase transitions including those \\ in modulated \( A'A''BX_{4} \) compounds.
In general  one may expect that our discussion of the polydispersive processes applies  to these systems too.
The polydispersive dielectric processes in \( K_{2}SeO_{4} \) and \( Rb_{2}ZnCl_{4} \) are 
related by some authors to non-linear systems of
discommensurations, where various kinds of modes are expected to be present.
Note: such an intuitive picture is compatible with our model calculations.
In \( \{N(CH_{3})_{4}\}_{2}ZnCl_{4} \) it would be interesting to perform
experiments with these mentioned crystals in which temperature and frequency dependence of the
effective relaxational frequency was observed and to
compare it with general expectations based on our theory.
While similar behaviour is observed also in the glasses of the RADP type,
their origin is more probable due to dynamics of defects and cluster walls, and not due to
dynamics of frozen-in incommensurate regions.

\section{Conclusions}

Interactions of the uniform mode with higher order modes due to incommensurately modulated
equilibrium state change the usual relaxation behaviour to a more complex one.
A multirelaxation character should be present even in the single-plane-wave limit.
Our model enables one to describe response of the incommensurable
modulated phases in order-disorder systems: while more detailed
mathematical  approach would predict known Cantor set type of the
fragmented energy spectrum, it would be difficult to verify such
spectrum and its consequences  in real materials. Our approach is
based on the idea of envelope description of this energy spectrum
and as such it certainly omits  some specific to Cantor sets
points of view. On the other hand approximations used here are
still of the predictive value: there are qualitatively new
predictions concerning response of the incommensurate phase
in order-disorder type of systems.

\end{document}